# Digital Twins, Internet of Things and Mobile Medicine: a Review of Current Platforms to Support Smart Healthcare

**Ivan Volkov**[1, [0000-0002-2084-8899],•], **Gleb Radchenko**[1, [0000-0002-7145-5630]],
**Andrey Tchernykh**[1,2,3 [0000-0001-5029-5212]]

[1] *South Ural State University, Chelyabinsk, Russia {volkovia, gleb.radchenko}@susu.ru*
[2] *CICESE Research Center, Ensenada, BC, México chernykh@cicese.mx*
[3] *Ivannikov Institute for System Programming, Moscow, Russia*
Received .......

**Abstract**—As the population grows, the need for a quality level of medical services grows correspondingly, so does the demand for information technology in medicine. The concept of "Smart Healthcare" offers many approaches aimed at solving the acute problems faced by modern healthcare. In this paper, we review the main problems of modern healthcare, analyze existing approaches and technologies in the areas of digital twins, the Internet of Things and mobile medicine, determine their effectiveness in solving the set problems, consider the technologies that are used to monitor and treat patients and propose the concept of the Smart Healthcare platform.

## 1. INTRODUCTION

Modern medicine faces multiple challenges related to the increasing need of the population for healthcare services. The demand for solutions from this area is driven by the growing volume of patient data, increasing technological capabilities, and the demand for fast and efficient healthcare processes and systems.

According to a 2010 study [36] the health information technology market was expected to grow from $99.6 billion in 2010 to $162.2 billion by 2015. Now, going back to the same source [35] we can see that in 2019 this market was at 187.6 billion and is expected to climb to 390.7 billion by 2024. Thus, we can see the exponential growth of this market in the previous decade, which is also projected for the next decade. The crisis caused by the COVID-19 epidemic only emphasizes this trend.

At the moment, specialists in digital twin technology, mobile medicine, and the Internet of Things are working on solving healthcare problems. All of these technologies are grouped in a field called «Smart Healthcare».

«Smart healthcare» is a system that addresses healthcare challenges with modern technologies and approaches to treatment and patient monitoring.

The article is organized as follows. Section 1 provides an overview of Smart Healthcare and digital twins. Section 2 focuses on popular approaches in Smart Healthcare. Section 3 provides an overview of mobile medicine in Smart Healthcare. In the last section, we propose a concept for the Smart Healthcare Platform

## 2. THE CONCEPT OF SMART HEALTHCARE

The concept of "Smart Healthcare" is part of the "Smart Planet" concept proposed by IBM in 2009. A smart planet is an intelligent infrastructure that uses sensors to acquire information, transmits that information through the Internet of Things (IoT), and processes it using supercomputers and cloud computing [15].

In this concept, smart healthcare is a healthcare system that uses technologies such as wearable devices, IoT and mobile Internet to provide dynamic access to information, connect people, materials and institutions relevant to healthcare, and then actively manage and respond to the needs of the medical ecosystem intelligently [26].

To understand why smart healthcare is a rapidly growing field in medicine, we review the problems that the concept of smart healthcare focuses on and the main technologies and approaches used to solve these problems.

## 2.1. *Problems of modern healthcare*

The key problems of modern healthcare are caused by such factors as [7]:

- an increase in the population and in life expectancy, which leads to an increase in the number of sick people who need the attention of doctors;
- the difficulty of monitoring patients' adherence to prescribed treatment;
- an increase in the number of elderly people [18] who require care and supervision;
- urbanization, which increases the chances of epidemics caused by the compact residence of large numbers of people. Such epidemics can lead to sharp jumps in the number of patients requiring medical care. One recent study, which was based on the current crisis caused by the COVID-19 epidemic, shows that high population density in cities causes accelerated growth of epidemics [4];
- a shortage of healthcare professionals who cannot maintain an adequate level of healthcare services to meet the growing needs of the population;
- an increasing cost of medical services, especially affecting patients with chronic diseases. For example, in the U.S. as of 2016, the cost of diabetes care was $245 million and has increased by 21% over 9 years [29].

These problems can be solved by applying modern technologies and approaches to the treatment of patients. For example, approaches in which the clinic and physicians are at the center of the process and patients do not have an active role in the treatment process are mostly used today. Allowing patients to take an active role in tracking and managing their health has the potential to help decentralize healthcare [20]that would reduce the burden currently placed on physicians and increase the effectiveness of the treatment provided.

## 2.2. *Digital twins in Smart Healthcare*

Application of the concept of "Digital Twins" today can be noted as one of the most striking trends in the digitalization of various industries. Emerged from the aerospace field, Digital Twin today is actively promoted to solve problems in industry, industry, management of complex systems (such as "Smart Cities"). In these areas, the concept of the Digital Twin is quite established. Among other things, it has allowed us to separate the concepts of a digital model, digital shadow and digital twin, the difference between which lies in the extent to which automation of data and control flows is provided, between a physical object (system) and its digital twin [13]. This approach to the definition can be called justified, because the use of the Internet of Things technologies, new approaches to the organization of data transfer, and providing control in the real-time lead to the fact that it is possible to provide synchronization of the state of the physical and digital twin in near-real-time.

The article [27] points out that developing a digital twin can be a very complex and therefore costly task, and can also increase the complexity of monitoring patient health in a hospital. Therefore, research related to digital twins needs to determine which data contribute most to the predictability of outcomes, how these outcomes can be evaluated, and how this approach can be cost-effectively integrated into healthcare. Ultimately, however, when properly implemented, digital twins have the potential to improve diagnostic and monitoring capabilities, improve therapy and patient well-being, reduce economic costs, and expand treatment options and patient options.

However, in the context of healthcare, the concept of a digital twin is not yet so clearly defined. This is due to the incredible complexity of the human being, an object of the physical world, for which a digital twin must be created. And if for components of industrial systems today there is a certain hope that application of modeling will allow providing the desired accuracy of results, but as for the human being, it is much more difficult to offer universal methods and approaches to modeling. Also, despite breakthroughs in the creation of new sensors and data collection methods, obtaining up-to-date data on key indicators of the human body in an operational mode is a difficult task, which can be implemented, most often, only in the laboratory clinical research settings.

## 3. SMART HEALTHCARE APPROACHES

Among the most popular approaches in Smart Healthcare are individualization, continuous health monitoring, telemedicine, and disease prevention.

### 3.1. *Individualization*

The concept of disease resulting from a causal factor is overly simplistic. Disease development is better described by understanding that each individual has an initial susceptibility to various diseases and may or may not get a disease as a result of environmental factors [24]. While previous approaches to medicine were static and focused on getting rid of the effects of a disease, the new approaches are dynamic and take into account that diseases develop over time (see Figure 1).

Individualization is an important factor here - each patient's course of illness, its cause, and the body's response to medications can be individualized. Because of this, the choice of treatment approach and patient monitoring can be affected by many factors, depending on the patient's body, their reactions to medications, their medical history, and their physiological parameters.

Thanks to modern technology, each person's health data can not only be collected and sto red but also analyzed, compared with data from other patients, and given feedback from the system in near real-time. With the proper use of technology, there will be no need for a lengthy analysis of a patient's response to the treatment applied to them, as the system will be able to select medications, dosages, and treatment plans almost immediately, focusing on multiple factors associated with a particular patient [34].

For example, the article [1] considers the possibility of using IBM Watson to help doctors work with patients with cancer. Clinical trials are at the heart of all medical advances in cancer prevention, detection, and treatment. However, just as no two people are alike, no two cancer pathways are alike. Now, coordinators analyze an average of 46 criteria to select patients for trials. This large range of data, which requires analysis and comparison with each patient, makes it difficult to accomplish this task without advanced analytical capabilities. With IBM Watson, a physician can automate the analysis process by providing patient-specific health information to the system. Watson analyzes patient data, comparing patient data to clinical trial databases, and offers the physician options for specific clinical trials appropriate for that patient.

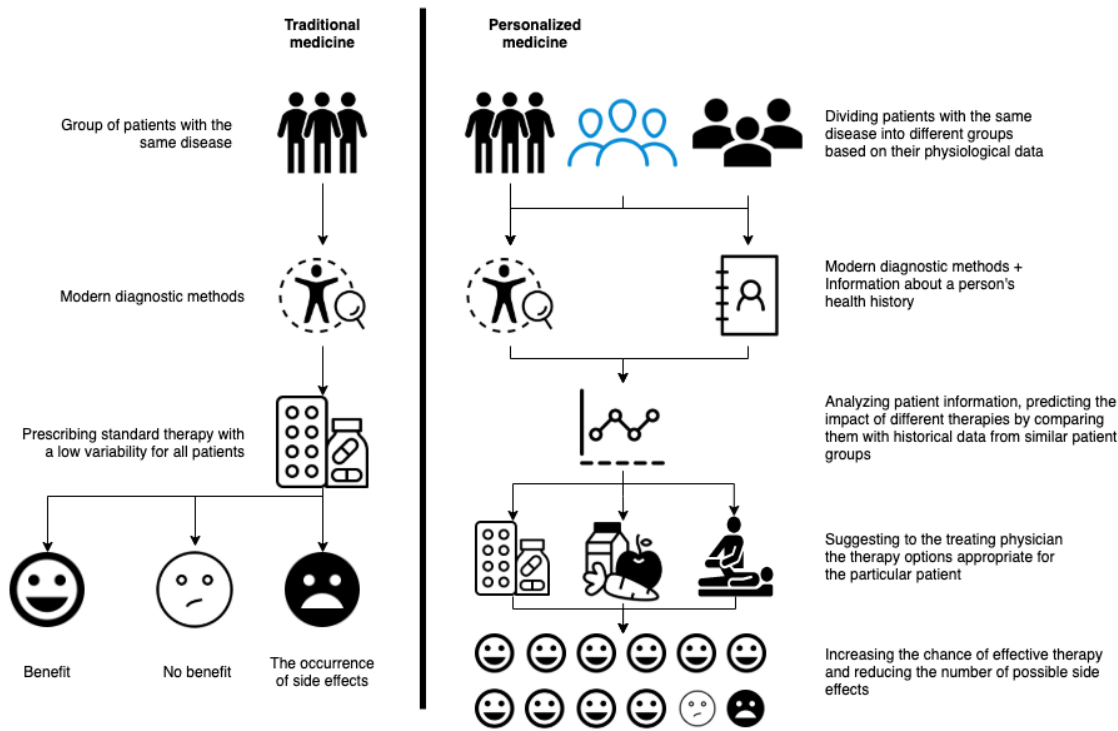

**Fig. 1**. Comparison of traditional and personalized medicine

However, an approach with this focus on individualization in healthcare has not begun to be considered long ago; it is being studied and developed thanks to advances in information technology and the emergence of smart devices in recent years. The authors of the article [12] point out that the effectiveness of individualization has not been sufficiently studied, and promises of a personalized approach to improve risk prediction, reduce costs, and improve public health for common diseases may in reality be less effective or simply unrealistic.

In the article [9] the authors also point out that data-driven decisions need to be better regulated because they raise partly unrealistic expectations and concerns. At the same time, they talk about the need to improve computational methods to provide measurable benefits in clinical practice.

### 3.2. Mobile medicine and continuous health monitoring

Mobile medicine is one of the key approaches today, providing a solution to the problems in the field of "Smart Health". Currently, there is a large amount of research work in the direction of Mobile health (mHealth - Mobile Medicine [8]), focused on methods of using mobile technology to continuously monitor and influence the patient's condition.

Mobile medicine includes not only technologies related to mobile applications, but also technologies of the Internet of Things, peripheral devices, computer vision, and telemedicine (see Figure 2).

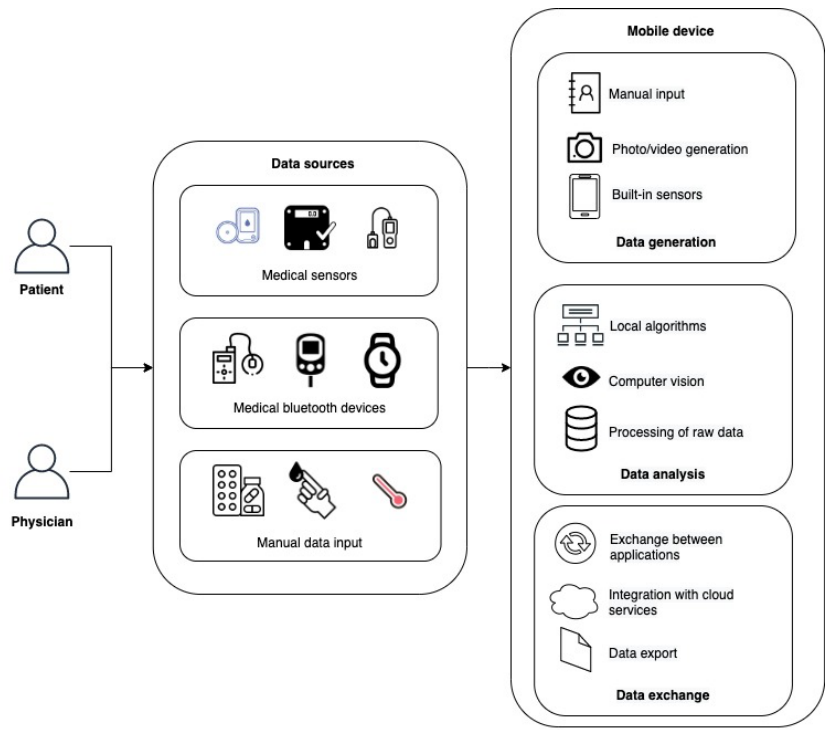

**Fig. 2**. Mobile medicine data sources and critical components

In 2016, doctors even saved a man's life with the help of mobile medicine. When a 42-year-old man was admitted to the hospital with a seizure, doctors looked at his Fitbit Charge HR fitness tracker's vitals, and it helped them make important decisions about the patient's future treatment path [21].

Mobile medicine encompasses many areas of development, among which the following can be noted [25]:

- *Wearable sensors* - bracelets, watches, headbands, patches, headphones, and clothing that provide passive and continuous monitoring of a person's biometric indicators;

- *on-chip laboratories* (complete analysis microsystems) are miniature instruments that allow one or more multistage (bio) chemical processes on a single chip with an area ranging from a few mm2 to several cm2 and using micro- or nanoscopic amounts of samples for sample preparation and reactions;

- *intelligent image analysis* - the high quality of smartphone cameras made it possible to use them for photometric diagnostics both with and without the use of additional devices (for example, to recognize an ear infection using an otoscope).

The growing number and variety of wearable devices, as well as the gradual decrease in their cost, make it easier to integrate them for monitoring patient health. Thus, mobile medicine is an important component for the realization of continuous health monitoring, which opens up new possibilities for doctors, patients, and researchers.

"Internet of Medical Things" - IoMT (Internet of Medical Things) is a separate category of the Internet of Things, the distinguishing feature of which is the use of sensors for monitoring and controlling the health of patients [16]. Such sensors enable the collection and processing of important biometric data about a person's health in real-time.

In the article [30] the main modern technology providing the growth of the sphere of health monitoring and the possibility of its mass implementation are distinguished by unobtrusive sensing and wearable devices (Unobtrusive sensing and wearable systems). The most commonly measured vital signs using such devices include: ECG, ballistocardiogram (BCG), heart rate, blood pressure (BP), blood oxygen saturation (SpO2), blood glucose levels, body temperature, posture and physical activity.

The main difference between these medical devices lies in the ability to integrate them into the user's daily life so that these devices will not only not disrupt the normal daily routine of the person, but rather give him the benefits of displaying detailed information about various aspects of his health on the screen of his mobile device.

This makes it possible not only to collect and store important data about a person's health over a long period autonomously and regularly, but also to analyze them, compare them with data from other patients, and help physicians make quick decisions based on current indicators.

In the article. [17] using this approach, the authors propose a solution to monitor the daily routines and behaviors of older adults to detect abnormal behavior without interfering with their lives. Such a solution could help in the care of elderly people with dementia or Alzheimer's disease who live alone.

### 3.3. *Telemedicine*

Advances in communication technology have given doctors and patients new opportunities to interact with each other. Telemedicine can manifest itself as a remote real-time video conference with a patient, as well as the ability to instantly exchange text or media information about a person's health. By doing so, the need for the physician and patient to be physically present in the same place can be eliminated in a multitude of human treatment processes, with many benefits.

Physicians can receive information from patients about their health much more quickly and conveniently and, accordingly, provide the patient with feedback about their health status and possible changes in the therapy they require as quickly as possible. These technologies are especially important when interacting with patients with chronic diseases who require constant monitoring and patients who are located in places remote from large medical institutions and who cannot quickly and safely get to a doctor for consultation or treatment adjustments.

The result of integrating telemedicine is lower costs for healthcare providers, increased patient satisfaction with the treatment process, and lower patient costs for chronic diseases (such as heart disease, diabetes, respiratory disease, or cancer) [2].

In this case, telemedicine also uses virtual reality (VR) technology. In 2009, there was a study [19]in which patients who had suffered a stroke were divided into two groups for rehabilitation. The first group underwent rehabilitation remotely under the control of a doctor with the help of virtual reality, the second group underwent rehabilitation in a local hospital. At the end of the study, no significant difference was found between the two groups, which shows the effectiveness of the use of virtual reality technology.

Studies have also been conducted on the use of augmented reality (AR) technology. In the article [28] the authors propose a technology that allows physicians in remote locations to be trained to perform complex medical procedures, such as ultrasound scans, without visual intervention. The mentor's hand gestures are captured using Leap Motion technology and virtually displayed in the space of the trainee's HoloLens glasses.

### 3.4. *Disease prevention*

Thanks to the widespread introduction of mobile devices and wearable sensors, users can know in advance if they are likely to become ill. A relevant example here would be the notification systems developed by Apple and Google about contacts with people infected with the COVID-19 virus. The presence of a cell phone in the vast majority of the population allows devices to anonymously collect and store information about the duration and number of a user's contacts with other people. This allows the system to anonymously notify all other people who have had contact with them of their possible risk of contracting the disease if one of the users tests positive for the disease as well [37].

In addition to detecting infection through contact tracing, modern technology makes it possible to analyze a person's health status and predict the development of diseases based on historical data on the person's biomedical indicators and their comparison with the historical data of all other users. Thus, the system can be trained not only to detect deviations from the standard human indicators but also to assume the cause of these changes and their possible further development. For example, in the article [6] the authors propose a machine learning algorithm, the implementation of which could allow predicting the risk of ischemic stroke in a patient being observed in a hospital with an accuracy of 94.8%. In another paper [22] propose an algorithm that the authors claim can use neural networks from voice recordings to determine whether a patient has Parkinson's disease with 100% accuracy.

## 4. MOBILE MEDICINE FOR SMART HEALTHCARE

Mobile devices are one of the main ways to easily collect data about a user's condition using mobile apps. At the moment, there are many apps to collect information about almost any disease by manually entering data by the user, or by reading data from sensors.

To understand how to effectively use mobile technology in healthcare, let's look at the existing problems in this area, options for the use of these technologies and existing commercial solutions

### 4.1. *Technology challenges in Mobile Medicine*

Biomedical sensors are now prevalent mainly as off-the-shelf devices that can be connected to an application on a mobile device via Bluetooth or Wi-Fi. Today, the data collected by such devices are limited to such types of indicators as heartbeat, physical activity during the day, and sleep quality.

At the moment, the level of implementation of wearable devices cannot provide a sufficient level of detail of human health information to provide accurate interpretations. There have been documented cases where the information collected from wearable devices has helped save a person [21] but more specific data collection is needed for a more detailed analysis of a person's health status.

The main problem is the mass creation, promotion and implementation of sensors that, on the one hand, will be convenient for the patient to wear at all times, and, on the other hand, will be affordable financially and functional enough to be able to reliably collect data and transmit them for processing to cloud services or mobile devices.

Also, each sensor requires the development of its application or web service, which will be responsible for data collection and processing. As a result, the data collected may end up on servers with different architectures without the ability to collect the resulting data in a single data center.

It should also be taken into account that wearable devices can generate a huge amount of information, several GB in one day from just one device [7]. Of course, not all of the received data needs to be transmitted over the network in a pure format, but the increase in the number of wearable devices and the information they generate leads to the need for an appropriate network architecture, which would be able to stably, reliably and timely transmit the data received to the interested parties.

### 4.2. Technology challenges in Mobile Medicine

The main problem of mobile applications in the healthcare industry is *the lack of mechanisms for coordinating heterogeneous information* collected by different applications from different sensors. Users install several apps on their devices, each of which stores and processes data in its format without the ability to exchange this information with each other.

Apple has the Apple Health app on its devices, with which Apple is trying to solve the problem of disparate data collection by providing an interface for applications to share data. The main problem is that it is a commercial product developed for the Apple ecosystem, which limits the use of this service by third parties.

Another problem is that developers do not need to implement interfaces to share data with HealthKit[1] as Apple Health currently does not provide the user with an in-depth analysis of the data stored in the app, and developers have no way to transfer data anywhere other than this app. And since most apps are now developed simultaneously for both iOS and Android systems, the lack of ability to exchange data between these platforms leads to the need to develop their web services, which will be responsible for storing and processing the data received.

Another important problem with the use of mobile apps is *the involvement of the patient in the treatment and monitoring of the disease*.

Gamification techniques of the disease monitoring process are used to solve this problem. A voluminous study reviewing 46 studies on gamification in healthcare was conducted in the article [23]According to the results of the study, most of the applications show a positive impact on the patient's health, helping in following the prescribed therapy, timely and better monitoring of the disease, and in general improving the attitude towards one's health and one's disease with increased motivation for therapy.

However, as the authors point out, most studies are built on small periods during which patients used the app. Because of this, it is difficult to say how effective the use of gamification is in the long-term treatment. Also, due consideration should be given to the patient's ability to cheat the system to gain more progress in the game, sacrificing the validity of the data entered into the app.

### 4.3. Using mobile medicine and Internet of Things technologies for patients with chronic diseases

Patients with chronic heart disease require continuous monitoring of their condition to

---

prevent critical situations before they occur. According to WHO statistics, about 230 million people have heart problems, up to 3 million people die from these problems each year [14]. Internet of Things technologies can greatly simplify and speed up the process of collecting data about the condition of the heart and transmit this data in real-time to the doctor and analyze it [16].

Large companies are now addressing this problem, for example, in 2019 Apple released the Apple Watch Series 4, which allows real-time ECGs of the heart and promptly notify users of abnormalities they have identified that need to be addressed [39].

Another disease that requires constant monitoring is diabetes mellitus, in which people need to keep track of their medical and nutritional records, recording these data about 10 times a day on average.

To solve the problem of collecting and analyzing diabetes data, some devices can automatically remember and synchronize data such as blood sugar levels.

Such devices can be divided into three groups:

- glucose meters, with the ability to synchronize data with your cell phone;
- systems for continuous monitoring of blood sugar levels with the display of this information on the device itself with the possibility of saving the data to a computer;
- a continuous sugar level monitoring system with the ability to read this data through the NFS chip in the device itself or in the cell phone.

A study [10] been conducted, the results of which suggest that these systems can greatly facilitate the process of collecting data on the user's sugar levels. This makes it possible to improve the patient's level of diabetes compensation and achieve more stable blood sugar levels. This is confirmed by the fact that in the patients who took part in the study, the average sugar level became closer to the target level than before the study began.

Some developments can not only collect information about blood sugar levels but also analyze them, providing the patient with information about current sugar levels and correcting them with special substances [31]. The technologies are still under development, but the presence of such projects indicates the presence of research in this area and the possibility of their further integration into future developments.

### 4.4. Commercial mobile medicine solutions

In a March 21, 2016 presentation, Apple announced two platforms it is developing: ResearchKit[1] and CareKit[2]

ResearchKit provides an API for collecting medical indicators of patients with various diseases. The purpose of the platform is to aggregate and analyze this data by specialized disease research institutions. This platform is already in the process of collecting information about Parkinson's disease, in which 9,520 people have participated and agreed to share their scores [3].

On the other hand, people's increasing desire to share their data, including medical data, can have unintended consequences and is extremely risky. Before the advent of mobile technology, it was not possible to collect daily information about a patient's condition so easily. There are risks that the mHealth platform may not be used for the purposes intended by its creators [11]. Therefore, the opportunity provided and the data obtained through it should be treated with the utmost caution.

The second platform, CareKit, is a framework that allows the creation of applications that can help users monitor their health. This platform allows simplifying the process of developing applications that help in the collection of users' health data for disease research.

Given the increasing amount of data collected through sensors and self-entry by the user, there is a need for a service that can aggregate all the data obtained in one place for subsequent analysis. As one solution to this problem, Apple has developed the Apple Health service, which allows users to store data obtained not only through sensors in Apple devices but also to synchronize data obtained by third-party applications through manual input or wearable devices.

---

[1] https://www.apple.com/ru/researchkit/

[2] https://developer.apple.com/carekit/

However, this solution imposes limitations on the choice of smartphones and wearable devices that users use and does not allow real-time access to the data received by attending physicians or relatives. One way forward should be to develop a comprehensive system that integrates all stakeholders at once:

- patients who would enter data about their health and lifestyle;
- medical personnel;
- medical researchers who organize studies that require large amounts of data from a variety of participants.

## 5. OVERVIEW OF MOBILE MEDICINE PLATFORMS

When selecting projects for analysis, we were guided by the following points:

- the project documentation is freely available;
- the project has real-world examples of use;
- the project does not focus on maintaining data on a specific disease, but rather on tracking the overall health of the individual as a whole

To better understand the differences in existing mobile medicine and smart healthcare platforms, we compared the currently available solutions based on the following criteria (see Table 1):

- Platform lifetime;
- Native AppStore application availability;

**Table 1.** Comparison of mobile medicine and smart healthcare platforms

| Platform / Features | Apple HealthKit | Google Fit | Microsoft HealthVault | HealthBox | Open mHealth |
|---|---|---|---|---|---|
| Platform lifetime | 2014 – present | 2014 – present | 2007 – 2019 | 2020 – present | 2011 – 2019 |
| Native AppStore application | + | + | + | – | – |
| Native Google Play application | – | + | + | – | – |
| iOS SDK | + | – | + | – | – |
| Android SDK | – | + | + | – | – |
| Server-side API | – | + | + | + | + |
| Web interface | – | – | + | – | – |
| Social sharing | + | – | – | – | – |
| Number of health data types | 160+ | 31+ | 80+ | 82 | 89 |
| Data types customization/ personalization | – | + | – | + | + |
| Data synchronization method | iCloud | Google Cloud | Microsoft Cloud | Own server | Own server |
| Data sharing with family members or physicians | +* | – | + | – | + |
| Self-hosted server requirement | – | – | – | + | + |
| Direct access to platform data from 3-rd party developers. | – | – | + | + | + |

\* Available starting from iOS 15

- Native Google Play application availability;
- iOS SDK availability;
- Android SDK availability;
- Server-side API - whether the platform allows you to receive, send and edit the data it contains through any data exchange protocol from another web application;
- Web interface - the ability to manage data via native web interface of the platform;
- Social sharing - whether the platform has the opportunity to share medical data with other people;
- Number of health data types monitored through the platform;
- The ability to customize/personalize data types – whether the platform allows you to configure data types not originally provided by the developer;
- Data synchronization method – technology used to synchronize the data;
- Data sharing with family members or physicians - at least one opportunity to share data other than creating a screenshot has been implemented;
- Self-hosted server requirement – whether you need to deploy your server to work with the platform;
- Direct access to platform data from 3-rd party developers – whether the developer needs to make an additional software to collect and analyze data from the platform.

Let's consider mobile medicine platforms that meet our defined parameters.

### 5.1. *Apple Health*

Apple Health was developed by Apple in 2014 and is still being maintained and developed today.

The main part of the platform is the Health mobile app, which is available for installation on Apple devices. The platform data is stored locally on the device and synchronized between the rest of the user's devices using iCloud technology.

The ability to read, modify and write data to local storage can be requested by any third-party apps available in the AppStore.

The platform has an extensive number of in-app inputs (more than 160), good documentation and support, providing easy integration with third-party apps. Most of the popular apps in the AppStore that allow you to monitor your health have implemented the HealthKit data synchronization feature.

Apple Health includes ResearchKit and CareKit frameworks, which provide researchers with the ability to build a ready-made health tracking and assessment application based on off-the-shelf modules. However, you still need to be a Swift or Objective C developer to use them.

Despite its great benefits, the main drawback of Apple Health is that it is isolated within the Apple ecosystem. Data from the platform can only be managed and shared directly through the mobile app. It is impossible to create a web service that would be able to manage user data in HealthKit anywhere other than iOS, regardless of the functionality of the mobile app. The lack of access to data via API also means that it is impossible to retrieve data on the Android operating system, which significantly narrows the range of users who can interact with the product implemented on this platform.

### 5.2. *Google Fit*

Google Fit[1] was developed by Google in 2014 and is still being maintained and developed today.

The main part of the platform is a web service that provides a single set of APIs for data management on the platform. Platform data is stored remotely on Google Cloud servers and can be synchronized locally on the device.

It has its mobile app in both Google Play and AppStore, allowing the user to interact directly with the platform by controlling a limited set of parameters (about 30 parameters in total) available in the platform.

---

[1] https://www.google.ru/fit/

For third-party developers, it is possible to manage user data from any device and application via the API, without having to contact the repository itself directly. To access user data, the third-party service must request the appropriate permissions from the user through their Google account. Connecting to the platform is quite easy by implementing the necessary API requests in your application.

However, despite the flexibility of interacting with the platform, the number of parameter types tracked in it is too small and the platform does not allow you to create your data types, which makes it impossible to use Google Fit as a universal health tracking platform.

### 5.3. *Microsoft HealthVault*

Microsoft HealthVault[1] - a project developed by Microsoft in 2007, but closed in 2019.

In terms of functionality, the platform had a fairly extensive set of advantages. There was an application in the AppStore, as well as in Google Play. In addition, the user could get access to the storage through the web interface. The data were all stored on the Microsoft Cloud service.

For third-party developers, the possibility to connect to the platform existed, but now it is difficult to say how easy it was to conduct integration since the documentation on the project is archived by the company. However, there is a freely available SDK for both iOS and Android that implements basic requests to the platform.

It was also possible to access the platform through an API, which allowed researchers to receive data from users directly, without the need to create their server.

The number and types of parameters in the project were quite extensive and allowed to keep most of the information about human health.

However, despite its flexible data management capabilities, the platform lacked any social component and any feedback on the patient's health status. This resulted in users not getting any practical benefit from using the platform, resulting in Microsoft shutting down the project in 2019 due to a competitive advantage [40].

### 5.4. *HealthBox*

HealthBox[2] - is an open-source project from independent developer "V0LT" published in 2020.

This project is not a finished product that can be downloaded from the AppStore or Google Play, or even opened in a browser. However, its goal coincides very well with the main goal of the smart healthcare platform, which is to centralize health data from different sources.

The developer of this project proposes a framework for a modular, centralized, and secure web service architecture that the user can use as-is or upgrade and then deploy as a Python-based web service to any device that allows it.

The obvious disadvantage of the project is its lack of prevalence and the lack of services that can integrate with it. At this stage, users are offered to create their adapters, which will transmit data to this service from other applications. However, this project was launched relatively recently, and development on it is continuing, so it is possible that in the future, this technology will find mass practical application.

### 5.5. *Open mHealth*

Open mHealth[3] - is an open-source project from independent developer Open mHealth, founded in 2011, but judging by the information on the project website, its development stopped in 2019. Despite this, this project is an excellent example of the Smart Healthcare platform approach.

This project, like HealthBox, is not a finished product and represents the basis for

---

creating a web service that allows you to collect and process medical data from different sources.

Although the project is developed by independent developers, during its existence it has had 9 use cases in such scenarios as an exchange of blood sugar data between the patient and the attending physician with the integration of data from 8 other services [33] and sharing and analyzing patient PTSD cases with data integration from 5 other services [5, 32].

The main difference between Open mHealth and HealthBox is a much more elaborate data schema and the principle of interaction with the platform, allowing not only to integrate data from different sources but also to process, visualize and send it to other services.

Despite the well-developed technical part, there are no cases of integration of this platform into mass projects, and all the examples given are either applied to software developers or are solutions developed for a particular patient. However, the approach and architecture offered by this platform can be taken as a basis when working on a unified Smart Healthcare platform.

## 6. SMART HEALTHCARE PLATFORM

Based on the analysis of existing solutions, we can say that a successful Healthcare Platform must support next requirements:

1- patients should receive practical benefit from the platform usage;
2- platform should contain social components;
3- data from the platform should be available for usage by any third-party developer without the neccessity to create custom-made web-service for data exchange;
4- plaform should allow patients to securely share their data with family members or physicians;
5- platform should be flexible in the terms of types of parameters available for monitoring;
6- platform should allow to create new parameters to configure the platform in a way that was not originally provided by the developer.

By looking at these requirements we can see that there are no solutions, that are matching all of them. Because of that, some of the platforms, like HealthVault and Open mHealth are now abandoned.

While other platforms are still active, they do not fully comply with the entire list of requirements, which makes it not possible to choose Apple Health or GoogleFit as a universal platform that could be used by any stakeholder for developing their universal Healthcare solution.

Based on that we propose a concept for the Smart Helathcare platform, supporting listed requirements.

### 6.1. Platform concept

At the moment, most Smart Healthcare solutions are developed independently of each other. For each problem, you either have to develop a new solution from scratch, or search for and integrate ready-made products.

Ideally, each of these parts should somehow interact with other parts of the system. But it is impossible to provide such support to independent developers and companies because each development solves its specific task in the way and with the means that were relevant at the time of writing the corresponding software.

A possible solution to this situation could be the creation of a platform that would offer medical institutions the ability to create the system they need based on pre-prepared modules that allow patients and physicians as well as other services to interact with the data stored in them. With such a platform, most of the time and financial costs required could be significantly reduced, since there would be no need to create a new development from scratch.

As a model of user interaction with this service, we propose to highlight the principle of a two-way marketplace, where the service provides users with the capabilities of a platform with a predetermined structure of data construction and interaction with them, with which companies can integrate to solve their internal tasks or to provide their services, and patients can receive services from companies that do not depend on each other but provide their services in one place and in a format agreed upon by the system in advance.

Analogous to this solution are platforms such as:

- *Coursera*, where independent instructors post their courses, available to all visitors to the site;
- *AppStore*, where independent developers post their apps that are available to all iOS users;
- *Alexa Skills Kit*, which allows independent developers to create their conversation scenarios with the voice assistant that can be used by Amazon Echo speaker owners.

The Smart Healthcare Platform should offer a similar use case, being a framework on which third-party developers and companies can host their services, available to all visitors to the platform.

### 6.2. *The main actors of the platform*

In the concept of smart healthcare, the patient-centered approach stands out as the main one. Within the field of information technology, it is possible to distinguish the following stakeholders.

1- *Patients.* Expect getting a wide range of medical services at an affordable price with personalized recommendations. In addition to getting a doctor's clinical diagnosis, they have the opportunity to gain more medical knowledge through digital platforms and connect with similar people for information such as disease symptoms, side effects, hospitalizations, medication information, clinical reports and developmental scenarios. The patient is the main source of information about his or her health through data generation on a mobile device.

2- *Healthcare providers.* The vast amount of data collected at various stages of patient diagnosis and treatment helps healthcare providers get a realistic picture of the proposed course of treatment. Health system data includes lab results, clinical notes, medical imaging data, and data from sensor devices. These data help improve public health surveillance and enable rapid response through effective analysis of disease patterns. Data from handheld devices help doctors track medication use, keeping track of a patient's health status at any given time.

3- *Clinical researchers.* The use of clinical data helps build predictive models for understanding biological and drug processes that contribute to high levels of efficacy in drug development. Analyzing medical data from a variety of sources helps clinical investigators measure drug development outcomes, even in small and rapid trials. Input from data from other participants in smart healthcare allows clinical organizations to assess and visualize their current situation to make strategic decisions.

### 6.3. *Platform processes*

Thus, to build a reliable and high-quality functioning solution in Smart Healthcare, it is necessary to implement the following processes:

- management of patients' physiological data;
- management of measurement data of the patient's medical parameters;
- data management via manual input;
- collection of data from sensors;
- system component management;
- data exchange between system components;
- Synchronization of data between the system components;
- Identification of abnormalities in the patient's state of health;
- visualization of the patient's health data;
- interpretation of the data obtained;
- delivery of alerts to the patient and the treating physician;

### 6.4. *Platform components*

Given the above factors, we have developed a component diagram for the implementation of the Smart Health Platform

In the proposed diagram (see Fig. 1), we divide the interaction with the platform into two main parts:

1- The client-oriented part, which is accessible to patients. On the client part, third-party developers can use the Smart Healthcare Platform SDK to implement their applications without having to deploy their own server;

2- The platform backend, which is responsible for processing data and deploying new applications to it. It provides all the stakeholders with access to the data on the platform through dedicated APIs. Third-party developers, meanwhile, can build other services on the platform, oriented for the usage by stakeholders.

This diagram shows an example of interaction with a single third-party application. In fact, any number of third-party applications can be connected to the platform and they will all exchange data in the same format and all data will be stored on the same platform.

## 7. CONCLUSIONS

As part of this work, we analyzed existing approaches in the field of "Smart Healthcare", which ensures their effectiveness in solving their tasks and the technologies that are used to monitor and treat patients.

The results of our analysis show that the field of Smart Healthcare is currently developing rapidly. The technologies of mobile medicine, digital twins and the Internet of Things can significantly improve the process of monitoring a person's health and positively influence their therapy.

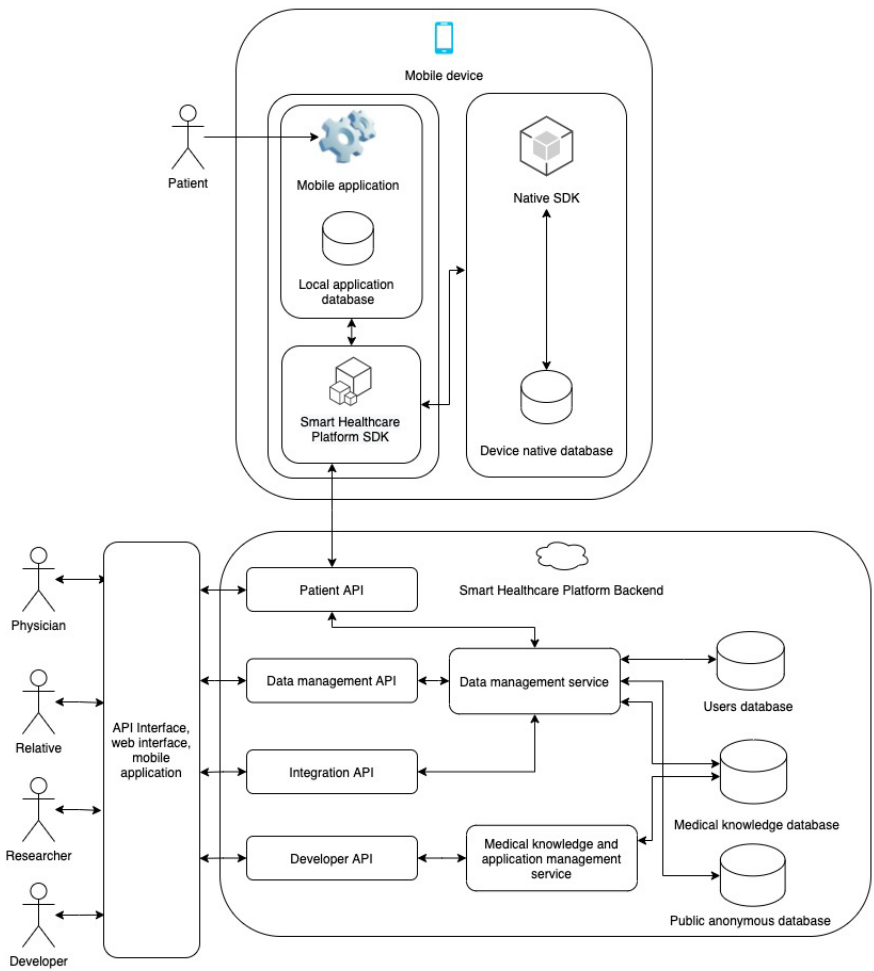

**Fig. 3.** Proposed Smart Healthcare Platform architecture

However, there is still insufficient data and research in these areas to assess the effectiveness of these technologies in their mass integration into healthcare structures. The creation of such technologies must necessarily be accompanied by an assessment of the practical and economic feasibility of their implementation.

It is also worth noting that currently all developments in this area are carried out independently of each other, and the various stakeholders have no opportunity to easily integrate medical services. The lack of standards and a single platform for structuring medical services complicates the process of integrating medical services, and there are more and more of them every year.

In the future, we plan to refine the concept of the Smart Healthcare Platform, define the requirements for such a platform and develop an architecture that would allow us to create a flexible, secure and reliable platform capable of adapting to the needs of a particular task.

**Acknowledgments.** Sections 3, 5 of the study was carried out with the financial support of the Russian Foundation for Basic Research and Chelyabinsk Oblast in the framework of scientific project № 20-47-740005, sections 2, 4, 6 were carried out with the financial support of the Ministry of Science and Higher Education of the Russian Federation (state task FENU-2020-0022).